\documentclass[aps,prb,twocolumn,superscriptaddress,showpacs]{revtex4}
\usepackage{amsmath,amssymb,amstext,graphicx,bm,bbm}
\usepackage{psfrag}
\makeatletter

\begin{document}
\title{ Spin Manipulation and Relaxation in Spin-Orbit Qubits}

\author{Massoud Borhani}
\affiliation{Department of Physics, University at Buffalo, SUNY, Buffalo, NY 14260-1500, USA}

\affiliation{Laboratory for Physical Sciences, 8050 Greenmead Drive, College Park, MD 20740, USA}

\author{Xuedong Hu}
\affiliation{Department of Physics, University at Buffalo, SUNY, Buffalo, NY 14260-1500, USA}

\date{\today}

\begin{abstract}
We derive a generalized form of the Electric Dipole Spin Resonance (EDSR) Hamiltonian in the presence of the spin-orbit interaction for single
spins in an elliptic quantum dot (QD) subject to an arbitrary (in both direction and magnitude) applied magnetic field. We predict a nonlinear
behavior of the Rabi frequency as a function of the magnetic field for sufficiently large Zeeman energies, and present a microscopic expression
for the anisotropic electron g-tensor. Similarly, an EDSR Hamiltonian is devised for two spins confined in a double quantum dot (DQD), where
coherent Rabi oscillations between the singlet and triplet states are induced by jittering the inter-dot distance at the resonance frequency.
Finally, we calculate two-electron-spin relaxation rates due to phonon emission, for both in-plane and perpendicular magnetic fields. Our
results have immediate applications to current EDSR experiments on nanowire QDs, g-factor optimization of confined carriers, and spin decay
measurements in DQD spin-orbit qubits.
\end{abstract}

\maketitle

An electron spin confined in a semiconductor quantum dot is a promising candidate for spin-based quantum information processing
\cite{LD,HansonRMP}. To implement reliable quantum gates and quantum registers, a complete theoretical knowledge of all single- and
two-spin-qubit decay channels is essential in order to mitigate the decoherence and relaxation of qubits. Spin-orbit interaction is one of the
most important mechanisms of spin mixing of different orbital states in semiconductor structures. When combined with phonons or charge
fluctuations, this spin mixture results in a variety of spin decay channels, although they have all turned out to be quite weak for confined
electrons \cite{KhaNaz,GKL,BGL}, leading to impressively long single-spin relaxation time ($T_1$) (of the order of seconds at 1 Tesla in GaAs
and minutes in Si:P) \cite{AmashaZumbuhl,Shankar}.  The dominant electron spin decoherence channel for GaAs QD and Si donors is pure dephasing due to
hyperfine interaction with the environmental nuclear spins \cite{BluhmNature,Cywinski}, with the corresponding coherence time ($T_2$) having
been pushed to few hundreds of microseconds in GaAs \cite{BluhmNature} and a few seconds in Si:P \cite{Shankar}.  This significant
separation of $T_1$ and $T_2$ time scales means that spin-orbit interaction can play a constructive role in qubits with a relatively strong
spin-orbit interaction.

Exploiting the orbital part of the electrons and holes to manipulate their spins (via the spin-orbit coupling) is becoming a common practice in
recent experimental and theoretical works on extended and confined electrons in heterostructures with strong spin-orbit interaction \cite{Zutic,
RashbaEfros, GBL,  HansonRMP, NowackScience, LeoSpinorbit,PettaEDSR}. GigaHertz manipulation of confined electrons in semiconductor QDs is
achievable, in the presence of an applied magnetic field, accompanied by an {\it ac electric} field \cite{GBL}.  For small Zeeman splitting
(compared to the orbital excitations), the Rabi frequency shows a linear dependence on both applied magnetic and electric fields. This
spin-electric-field coupling was observed in transport experiments on lateral GaAs QDs and InAs nanowires in the spin-blockade regime
\cite{NowackScience,LeoSpinorbit,PettaEDSR}, where the Rabi spin-flip time of 10 ns has been reported. Such a fast spin rotation is a key step
towards realizing single- and two-qubit gates for localized spin qubits \cite{LD,HansonRMP}.

In this paper, we first present the general EDSR Hamiltonian for a single electron spin confined in an {\it elliptic} 2D QD, in the presence of
an arbitrarily large magnetic field.  The study of elliptic confinement is crucial in nano-wire QDs, where the electron orbital wave function is
strongly squeezed in two directions \cite{Samuelson,Ensslin}. In contrast to previous works \cite{RashbaEfros,GBL}, we show that the Rabi
frequency exhibits a pronounced non-linear behavior as a function of the applied magnetic field for sufficiently large QDs and Zeeman energies.
We also calculate the second-order corrections of the spin-orbit coupling to the EDSR Hamiltonian, and provide an analytical microscopic
expression for the g-factor renormalization.  The ability to engineer the g-factor of a confined electron is essential for individual addressing
of spin-orbit qubits via electric pulses \cite{LeoSpinorbit,PettaEDSR}.  In addition, we propose a scheme to perform coherent rotation in the
singlet-triple subspace of two electron spins confined in a double QD. Again, Zeeman energy and the spin-orbit interaction play the central role
in the electrical manipulation of the two-spin states, where the corresponding Rabi frequency is linear, in the leading order, in magnetic field
and the spin-oribt coupling. The proposed two-electron EDSR offers an alternative way to implement coherent swap gates in coupled spin-orbit
qubits, without a magnetic field gradient. Finally, we calculate the phonon-induced spin relaxation time of the two-spin system and find that
the resulting time scales are relatively long, which justifies the efficiency of our two-spin EDSR scheme.

{\it The effective EDSR Hamiltonian for an elliptic dot.} Consider a single electron in an elliptic 2D quantum dot, with confining frequencies
$\omega_x$ and $\omega_y$, subject to an applied magnetic field ${\bm B} = B (\sin \theta \cos \phi, \sin \theta \sin \phi, \cos \theta)$.  The
Hamiltonian for the electron is
\begin{eqnarray}
H &=&  H_d + H_Z + H_{so}, \label{H}\\
H_d &=& \frac{p^2}{2 m} + \frac{1}{2}m (\omega_x^2 x^2 +\omega_y^2 y^2),\\
H_Z&=&\frac{1}{2}g\mu_B{{\bm B}}\cdot\mbox{\boldmath $\sigma$}
=\frac{1}{2}E_Z{{\bm n}}\cdot\mbox{\boldmath $\sigma$},\\
H_{so} &=& \mbox{\boldmath$\cal{K}$} \cdot \mbox{\boldmath$\sigma$}, \;\;\;\;\;\;\;\;\;\;\;
\mbox{\boldmath$\cal{K}$} \equiv \frac{\hbar}{m}(p_y/\lambda_-,p_x/\lambda_+,0),\;\;\;\;\;\;
\end{eqnarray}
where $m$ is the effective mass of the electron, ${\bm p} = -i \hbar \partial/\partial{\bm r} + (e/c)\bm{A}(\bm{r})$ the electron kinetic
momentum in two dimensions, and $c$ the speed of light in vacuum.  $\lambda_\pm = {\hbar}/{m (\beta \pm \alpha)}$ are the effective spin-orbit
lengths in terms of the linear Rashba ($\alpha$) and Dresselhaus ($\beta$) coupling constants \cite{Rashba, Dress}. We restrict our
consideration to quantum dots with strong confinement along one axis ([001]), such as, e.g., quantum dots defined in a two-dimensional electron
gas (2DEG). With a uniform ${\bm B}$ field, the vector potential is ${\bm A}({\bm r})=B_z(-y/2,x/2,0)$ in the symmetric gauge.  With the motion
along $z$ strongly quantized (the 2DEG thickness ${\cal D} \ll \sqrt{\hbar c/e B_{x(y)}}$), the in-plane components $B_x$ and $B_y$ are not
included in ${\bm A}({\bm r})$.  The magnetic field also induces a Zeeman splitting $E_Z=g\mu_BB$, with a spin quantization axis ${\bm n}={\bm
B}/B$. Note that the linear-in-momentum spin-orbit interaction is given in a coordinate system which is rotated by $\pi/4$ along the $z$ axis
with respect to the crystallographic axes of the crystal and it is the sum of both Rashba and Dresselhaus terms.

\begin{figure}
\begin{center}
\includegraphics[angle=0,width=0.5\textwidth]{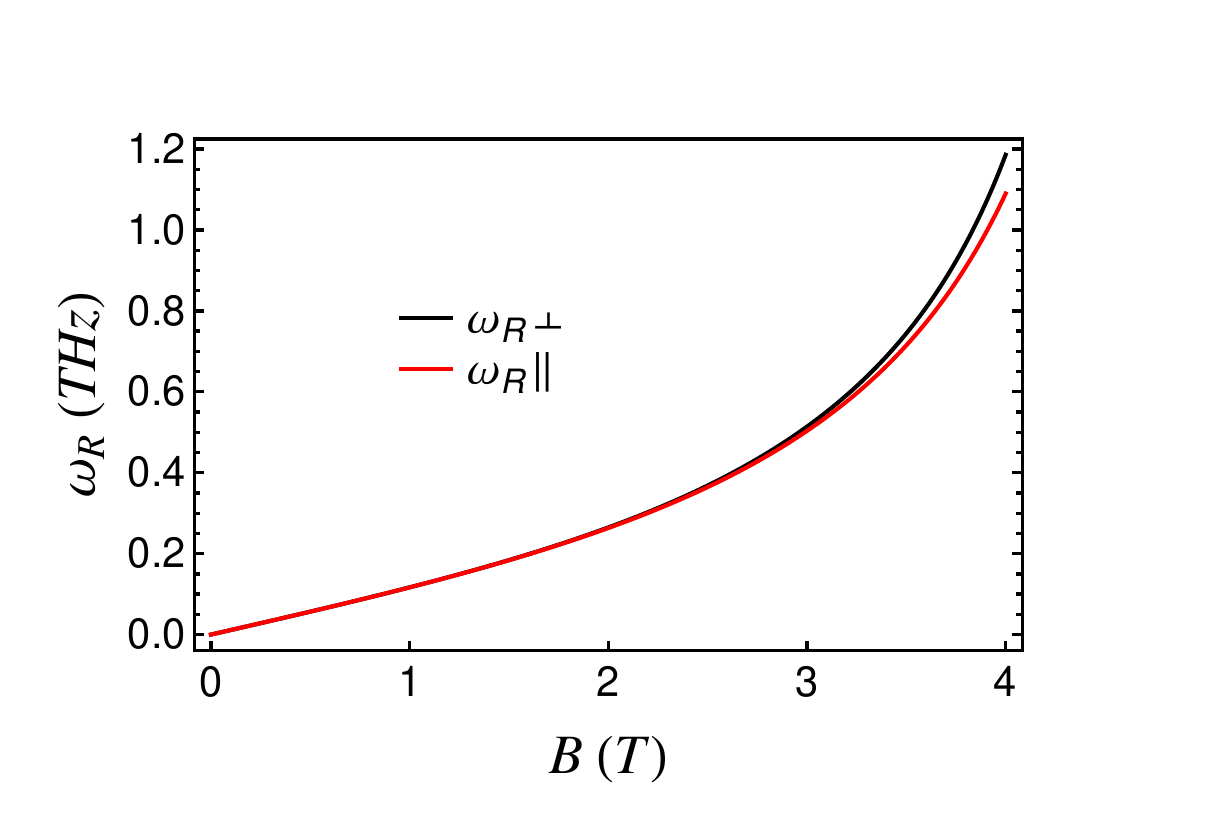}
\caption{\small (color online).  The Rabi frequency versus the magnetic field for an electrically driven single spin in a typical InAs nanowire
QD \cite{Samuelson}: $\omega_x = 3$ meV, $\omega_y = 40$ meV, and $\lambda_{+} \sim 100$ nm, for both perpendicular ($\omega_{R\perp}$) and
in-plane ($\omega_{R\parallel}$) magnetic fields. The electric field is in-plane and along the $x$ axis. The non-linear dependence shows up for
$B$ fields above $2$ T, although this threshold can be substantially reduced for larger dots and/or for materials with stronger spin-orbit
interaction.  } \label{Rabi}
\end{center}
\end{figure}
In order to diagonalize the Hamiltonian in Eq. (\ref{H}), we invoke the Schrieffer-Wolff transformation $H \rightarrow e^{-{\cal S}} H e^{\cal
S}$ \cite{SW,BGL}, and remove the spin-oribit interaction in the leading order by imposing $ [H_d + H_Z, {\cal S} ] = H_{so}$. The resulting
${\cal S}$ matrix is presented here to all orders in the Zeeman interaction, but in the first order in spin-orbit coupling: ${\cal S}={\cal
S}^{(0)} + {\cal S}^{(1)} + {\cal S}^{(2)}$,
\begin{eqnarray}
{\cal S}^{(0)} &=& i \mbox{\boldmath$\xi$}\cdot\mbox{\boldmath$\sigma$}, \label{Smatrix0}\\
{\cal S}^{(1)} &=&  i\left[\mbox{\boldmath$n$} \times \mbox{\boldmath$\eta$} \right]\cdot\mbox{\boldmath$\sigma$}, \label{Smatrix1}\\
{\cal S}^{(2)} &=& i \left[\mbox{\boldmath$n$}\times (\mbox{\boldmath$n$} \times \mbox{\boldmath$\gamma$} )
\right]\cdot\mbox{\boldmath$\sigma$}, \;\;\;\;\;\;\;  \label{Smatrix2}\\
\mbox{\boldmath$\xi$} &=& \left( y/\lambda_-,\, x/\lambda_+,\, 0 \right), \;\;\;\;\;\;\; \label{xi} \nonumber\\
\mbox{\boldmath$\eta$} &=& \left( \alpha_1 p_{y} + \alpha_2 x,\, \beta_1 p_{x} +\beta_2 y,\, 0 \right), \;\;\;\;\;\;\; \label{eta} \nonumber\\
\mbox{\boldmath$\gamma$} &=& \left( \tilde\alpha_1 p_{x} + \tilde\alpha_2  y  ,\, \tilde\beta_1 p_{y} + \tilde\beta_2 x  ,\, 0 \right).
\;\;\;\;\;\;\; \label{gamma} \nonumber
\end{eqnarray}
The coefficients $\{\alpha_i ,\beta_i,\tilde\alpha_i,\tilde\beta_i\}$ are given in Table I.  Due to the spin-orbit interaction, the spin and
orbital part of the rotated states are strongly coupled to each other, which opens up the possibility of spin manipulation via driving by an
{\it ac} electric field $ V(\mbox{\boldmath$r$},t) = e \mbox{\boldmath$E$} \cdot \mbox{\boldmath$r$}$ (within the dipole approximation).  The
effective Hamiltonian, $ H_{EDSR} = H_Z + \langle \psi_0 | [S,V({\mbox{\boldmath$r$}},t)] | \psi_0 \rangle $ for the spin sector of an electron
confined in the ground orbital state of an elliptic dot, is obtained by averaging over the orbital ground state 
\begin{eqnarray}
  H_{EDSR} &=& H_Z  +\frac{1}{2}\mbox{\boldmath$h$} (t)\cdot \mbox{\boldmath$\sigma$}, \label{H_EDSR}\\
  \mbox{\boldmath$h$} (t) &=& 2 \mbox{\boldmath$n$} \times (\mbox{\boldmath${\cal F}$} + \mbox{\boldmath${\cal G}$}),\\
  \mbox{\boldmath${\cal F}$} &=& e \hbar \; (\alpha_1 E_y, \beta_1 E_x, 0), \nonumber\\
  \mbox{\boldmath${\cal G}$} &=& e \hbar \; (-n_z  \tilde\beta_1 E_y, n_z \tilde\alpha_1 E_x, n_x
  \tilde\beta_1 E_y - n_y \tilde\alpha_1 E_x ). \;\;\;\;
  \nonumber
 \end{eqnarray}
\begin{widetext}
\begin{table*}
\caption {The $\{\alpha_i, \tilde\alpha_i \}$ coefficients of the transformation matrix $ {\cal S}$ in Eqs.~(\ref{Smatrix0}-\ref{Smatrix2}) as a
function of the Zeeman interaction, the spin-orbit lengths, the cyclotron frequency $\omega_c = e B_z /m c$, and  the lateral confining
frequencies. Note that ${\cal M}$ is a common factor which appears in all coefficients. Similarly, $\{\beta_i, \tilde\beta_i \}$ are given by
the following substitutions: $\beta_1 = \alpha_1 (\omega_x \rightarrow \omega_y, \lambda_- \rightarrow \lambda_+)$, $\beta_2 = - \alpha_2
(\omega_x \rightarrow \omega_y, \lambda_- \rightarrow \lambda_+)$, $\tilde\beta_1 = - \tilde\alpha_1 (\lambda_- \rightarrow \lambda_+)$, and
$\tilde\beta_2 = \tilde\alpha_2 (\omega_x \rightarrow \omega_y, \lambda_- \rightarrow \lambda_+)$; For example, $\beta_2= \hbar E_Z
(\hbar\omega_c)(\hbar\omega_y)^2 / m {\cal M} \lambda_+$.}
\begin{tabular}
{|c||c|c|c|c|}
\hline
$\cal M$ & $ m  {\cal M} \lambda_-  \alpha_1 / \hbar $ & $ {\cal M} \lambda_-  \alpha_2  $&
$ m {\cal M} \lambda_- \tilde\alpha_1/ \hbar $& $ {\cal M} \lambda_- \tilde\alpha_2$ \\
\hline
$E_Z^2 (\hbar\omega_c)^2 - [(\hbar\omega_x)^2 - E_Z^2][(\hbar\omega_y)^2 - E_Z^2]$ & $E_Z [(\hbar\omega_x)^2 - E_Z^2 ]$
& $ - E_Z (\hbar\omega_c)(\hbar\omega_x)^2$ & $ E_Z^2 (\hbar\omega_c)$ & $ E_Z^2 [ (\hbar\omega_x)^2 +  (\hbar\omega_c)^2- E_Z^2 ] $ \\
\hline
\end{tabular}
\end{table*}
\end{widetext}

We note that the effective {\it ac} {\it magnetic} field $\mbox{\boldmath$h$}(t)$ vanishes in the absence of the applied magnetic field, and is
always perpendicular to $\mbox{\boldmath$B$}$ . This is a distinct feature of the linear spin-orbit interaction in the leading order. Moreover,
for an in-plane magnetic field, the vector $\mbox{\boldmath${\cal G}$}$ is identically zero because $\omega_c = 0$ (see Table I).

In general, Rabi frequency is strongly dependent on the directions of the magnetic and the $ac$ electric fields, the confinement frequencies,
and the orientation of the QD with respect to the crystallographic axes. For small Zeeman energies ($E_Z \ll \hbar \omega_x, \hbar \omega_y$),
the Rabi frequency is linear in $E_Z$ \cite{GBL}.  In Fig.~\ref{Rabi}, we have plotted the Rabi frequency $\omega_R= max \, |\mbox{\boldmath$h$}
(t)|/\hbar$, between the Zeeman sublevels, as a function of the applied $B$ field. A clear non-linear behavior arises for magnetic fields above
2 T. To obtain these results, we used parameters for a typical InAs nanowire QD. This nonlinearity is a direct consequence of the {\it exact}
treatment of the Zeeman interaction in calculating the Schrieffer-Wolff matrix ${\cal S}$.

The above consideration is restricted to a two-dimensional QD where the effective spin-orbit interaction is linear in momentum. In reality, all
QDs have finite thickness ${\cal D}$ of the order of a few nanometers. Therefore, the {\it cubic} Dresselhaus term $H_{so}^{3D} \sim \frac{\beta
{\cal D}^2}{\hbar^2} \left(p_yp_xp_y\sigma_x-p_xp_yp_x\sigma_y\right)$ can also, in principle, induce spin transitions \cite{Dress}, though
in the leading order it is ineffective in the presence of harmonic confinements and in-plane magnetic fields \cite{GBL}.

For spin manipulations in QDs, a comprehensive knowledge of the electron g-tensor is crucial \cite{LeoSpinorbit,PettaEDSR}.  We have calculated
the second-order corrections of the linear spin-orbit interaction to the effective Hamiltonian, Eq.~(\ref {H_EDSR}), which lead to the
renormalization of the g-factor \cite{GBL,Flatte}.  For an in-plane magnetic field along the $x$ axis, the renormalized g-tensor $\tilde g_{ij}$
of an electron in the ground state of the harmonic potential reads (results for an arbitrary $B$ field direction is presented in appendix A)
\begin{eqnarray}
 \tilde g_{xx} &=& g_{xx} \left(1- \frac{\hbar^3 \omega_x}{ m \lambda_+^2 [(\hbar\omega_x)^2 - E_Z^2]}\right), \nonumber \\
 \tilde g_{yy} &=& g_{yy} = \tilde g_{zz} = g_{zz} = g_{xx}= g. \;\;\;\;\;\;\;
\end{eqnarray}
This renormalization of the g-factor can be used to address nearby spins selectively, and has immediate applications in current EDSR experiments
on confined carriers \cite{LeoSpinorbit,PettaEDSR}.

\begin{figure}
\begin{center}
\includegraphics[angle=0,width=0.5\textwidth]{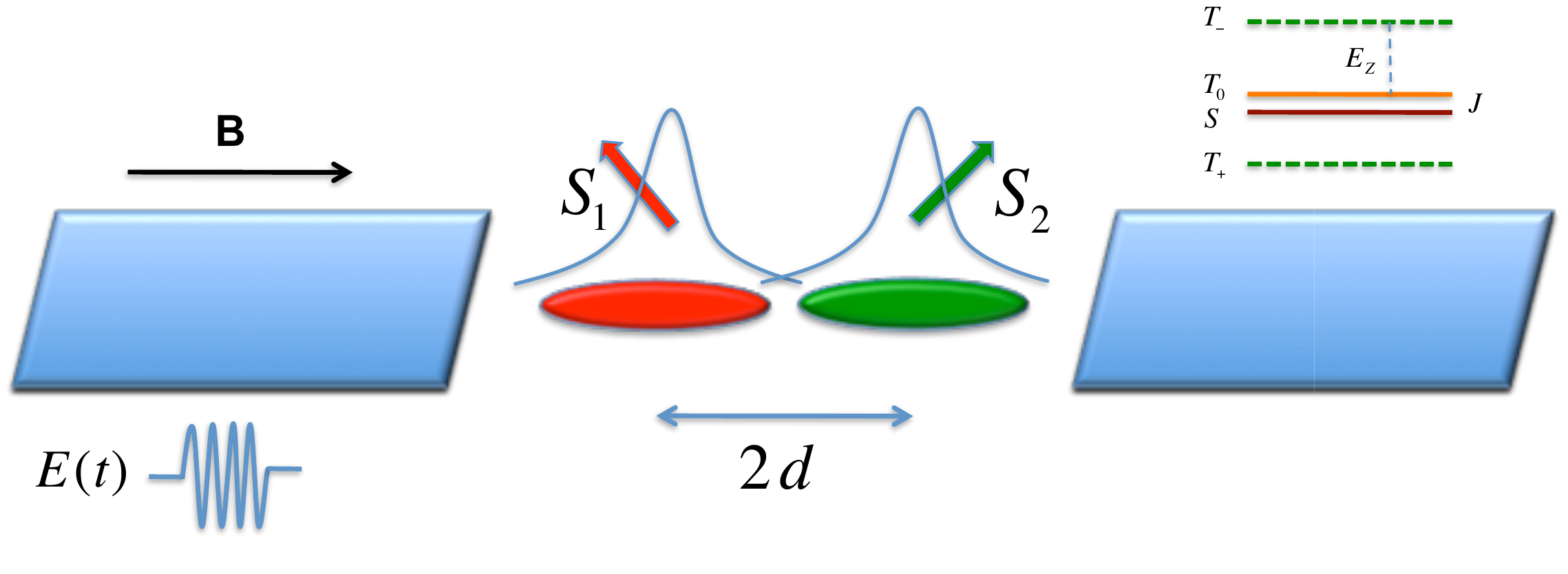}
\caption{\small (color online).  Schematic of a singlet-triplet DQD spin qubit \cite{PettaScience} in the presence of an in-plane magnetic field
and an {\it ac} electric field. The distance between the center of the double elliptic dots is $2d$ and the driving electric field is applied
along the DQD in order to induce maximum electric dipole moment. In the upper right, the corresponding energy levels of two-spins has been shown
for the regime $J \ll E_Z$.} \label{DD}
\end{center}
\end{figure}

{\it Two-electron spin Hamiltonian.} We now consider two electrons confined in a DQD separated by $2 d$, as shown in Fig.~\ref{DD}, and the
vector $\mbox{\boldmath$d$}$ denotes the orientation of the DQD with respect to the crystallographic axes. This setup has been the cornerstone
of recent experimental works on spin-selective transport, where the lifetime of different spin states are examined through the spin-blockade
effect \cite{OnoScience,NowackScience,LeoSpinorbit,PettaEDSR,PettaScience}.  In the absence of the spin-orbit interaction, the low energy
subspace of the two-electron Hilbert space consists of the following four states
\begin{eqnarray}
\{ \Phi_i \}_{i=1,...,4}  = \{ \Psi_+ S, \Psi_- T_+, \Psi_- T_0, \Psi_- T_- \}
\end{eqnarray}
where $S$ and $T_{\pm,0}$ refer to the spin singlet and triplet states, respectively, and $ \Psi_{\pm}$ are their corresponding symmetric and
anti-symmetric orbital wavefunctions.  By including the spin-orbit coupling, the singlet and triplet states are coupled provided that an
external magnetic field is applied.  For small magnetic fields, the effective Hamiltonian for the two spin confined in a double dot is given by
\cite{StanoPRLPRB}
\begin{eqnarray}
H_{ex} &=& (J/4)  \; \mbox{\boldmath$\sigma$}_1 \cdot \mbox{\boldmath$\sigma$}_2
+ \mu  \mbox{\boldmath$B$}  \cdot ( \mbox{\boldmath$\sigma$}_1 +  \mbox{\boldmath$\sigma$}_2) \nonumber\;\;\;\;\;\\
&&+ \; \mbox{\boldmath$a$} \cdot ( \mbox{\boldmath$\sigma$}_1 -  \mbox{\boldmath$\sigma$}_2) + \mbox{\boldmath$b$} \cdot (
\mbox{\boldmath$\sigma$}_1 \times  \mbox{\boldmath$\sigma$}_2) \label{PeterH}
\end{eqnarray}
where $\mu \equiv g \mu_B /2$ and $J$ is the singlet-triplet exchange splitting. Here we keep only the leading order terms in spin-orbit
coupling and neglect the second and higher order terms. The only linear terms in spin-orbit (and Zeeman) interaction are the following vectors
\begin{eqnarray}
\mbox{\boldmath$a$} &\equiv& -\mu \mbox{\boldmath$B$} \times {\text {Re}} \; \langle \Psi_+ |  \mbox{\boldmath$\xi$} | \Psi_- \rangle, \\
\mbox{\boldmath$b$}  &\equiv& -\mu \mbox{\boldmath$B$} \times {\text {Im}} \; \langle \Psi_+ |  \mbox{\boldmath$\xi$} | \Psi_- \rangle,
\end{eqnarray}
where $a=| \mbox{\boldmath$a$} |$ and $b=| \mbox{\boldmath$b$} |$ denote the strength of the spin-orbit matrix elements.  Note that the
Hamiltonian in Eq.~(\ref{PeterH}) has the rotated spin-orbit states $\{ U^{0} \Phi_i \}$, where $U^{0}= {\text {exp}} [-{\cal S}^{(0)}] $
\cite{StanoPRLPRB} as the basis.  For an in-plane magnetic field, $S$ and $T_0$ are coupled via the $z$ component of the vector
$\mbox{\boldmath$a$}$ (there is no coupling, in the leading order, between these two states for a perpendicular magnetic field).  Specifically,
$a_z = \frac{\mu B d}{2 \lambda_+}$ (in the Heitler-London approximation), provided that $\mbox{\boldmath$B$}$ and $\mbox{\boldmath$d$}$ are
along the $x$ axis. Rabi Oscillations between the singlet and the triplet states have already been observed in DQDs using the hyperfine field
gradient \cite{PettaScience}. When spin-orbit interaction is included in the consideration, a time-dependent $a_z$ can also induce coherent
oscillations between $S$ and $T_0$ states, providing an alternative way to manipulate the singlet and triplet states. For example, a sinusoidal
electric field pulse can change the inter-dot distance periodically to form a {\it breathing} DQD, $d = d_0 + \delta \sin (\omega t)$. The
corresponding Rabi frequency is then given by $\omega_R = \frac{\mu B \delta}{2 \hbar \lambda_+}$, where $\delta \ll d_0$ is the amplitude of
the breathing DQD, and $\delta$ is linearly proportional to the applied $ac$ electric field $E$. For InAs QDs, where the g-factor $g\approx 10$
and $\lambda_+  \approx 100$ nm, a $\delta \approx 1$ nm at $B=1$ T would allow a $\pi/2$ pulse on the order of 1 ns.

The exchange coupling between the two spins also oscillates as we periodically change their relative distance. Geometrically, this leads to a
breathing Bloch sphere which beats at the driving frequency. However, if the amplitude of modulations is reasonably small, they yield small
deviations $\delta J$ from the equilibrium value $J$ and the coherent Rabi oscillations should still be observable in the leading order of the
spin-orbit interaction \cite{GBL}.
\begin{figure}
\begin{center}
\includegraphics[angle=0,width=0.45\textwidth]{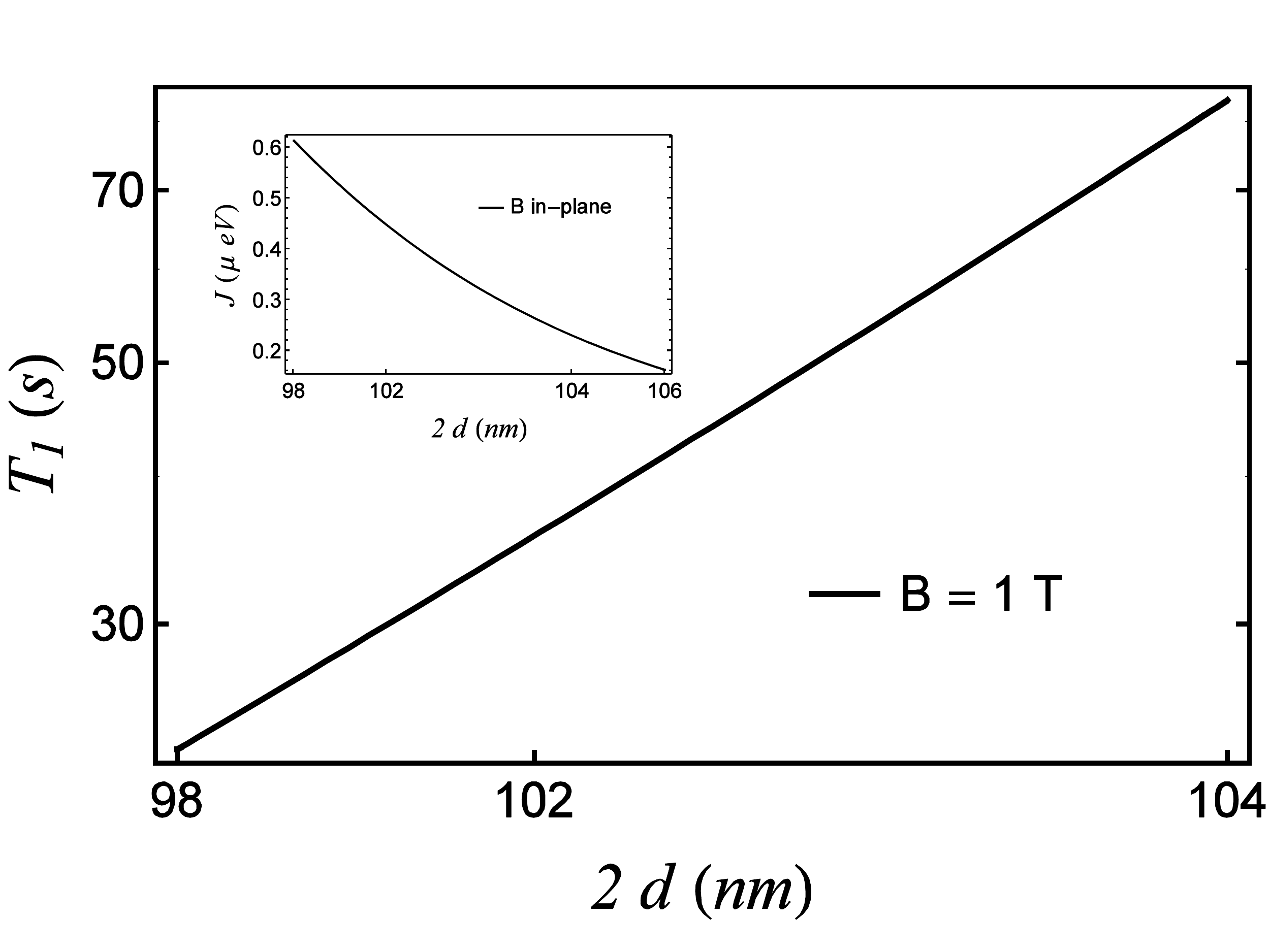}
\caption{\small The logarithmic plot of the triplet-singlet ($T_0$-$S$) relaxation time, for a DQD along the $x$ axis, as a function of the
distance between the two QDs. The applied magnetic field $B=1 T$ ( in this case, the rate has quadratic dependence on the field). The inset
shows the corresponding dependence of $J$ on $2d$, calculated in the Heitler-London approximation. The material parameters are chosen for GaAs
quantum dots with $\lambda_{+} \sim 5 \; \mu$m,  $l= 22$ nm, $e_{14} = 1.4 \times 10^9$ V/m, $\kappa=13.1$, $\rho = 5.3 \times 10^3$ kg/m$^3$,
and $s_t = 2.5 \times 10^3$ m/s.  } \label{Relaxation}
\end{center}
\end{figure}

{\it Singlet-triplet relaxation rates.}  Due to the spin mixing via the spin-orbit interaction, phonon emission can lead to relaxation (and/or
leakage out) of the two-spin singlet and triplet states.  In III-V semiconductors like GaAs or InAs, piezoelectric coupling to acoustic phonons
is the dominant relaxation channel \cite{Mahan,GKL}.  With an in-plane magnetic field (in which regime $S$-$T_+$ and $T_0$-$T_+$ do not couple
in the leading order) and in the leading order of the spin-orbit coupling, the triplet-singlet relaxation rate is given by
\begin{eqnarray}
\Gamma_{T_0 \rightarrow  S} &=& \frac{4 (e e_{14})^2}{\pi \kappa^2 \rho \hbar^2 s_t^3}  \cdot (n_{q_0}+1)
 \; e^ {-4d^2/l^2} \cdot \frac{a_z^2}{J} \; f(u) , \label{rate-in}\;\;\;\;\;\;\;\;\\
 a_z &=& \frac{\mu}{2} \left( \frac{B_x d_x}{\lambda_+} - \frac{B_y d_y}{\lambda_-} \right) \\
 f(u)&=& \frac{21u-2u^3+[4u^2(u^2-3)-21]D(u)}{2u^5} \nonumber  \\
 D(u) &=& e^{-u^2}\int_0^u e^{v^2} dv, \;\;\;\;\;\;\; u \equiv \frac{lJ}{\sqrt {2} \hbar s_t},
 \nonumber
 \end{eqnarray}
where $e_{14}$ and $\kappa$ are the piezoelectric and the dielectric constants of the material, respectively, $\rho$ is the density of the
sample, $s_t$ is the speed of transverse acoustic phonons, and $l$ is the size of each quantum dot.  This relaxation rate is valid when $a_z \ll
J$.  Moreover, $a_z$ has a geometrical structure which enables us to mitigate this relaxation channel by carefully choosing the direction of the
magnetic field with respect to the crystallographic axes and the DQD orientation. In Fig.~\ref{Relaxation}, we have plotted the relaxation rate
in Eq.~(\ref{rate-in}), together with the exchange splitting (within the Heitler-London approximation), as a function of the DQD separation $2d$
in the presence of an applied in-plane magnetic field of 1 T. The resulting time scales are relatively long for small Zeeman energies (below 1
Tesla).  Therefore, we conclude that in this regime single-spin and/or hyperfine-induced triplet-singlet relaxations dominate the spin
relaxation processes, while dephasing is probably dominated by hyperfine interaction \cite{Cywinski}, charge noise \cite{Hu_PRL06}, and phonons
\cite{Hu_PRB11}. Note that while the Hamiltonian in Eq.~(\ref{PeterH}) is quite generic for any inter-dot coupling, the analytical results
presented here are for a relatively narrow inter-dot distance regime, because our analytical calculation is bound from the small $d$ by the
validity of our Heitler-London model to calculate $J$ and $a_z$, and from the large $d$ by the requirement that singlet and triplet states are
close to the two-electron eigenstates.

For a perpendicular magnetic field, the only states which are coupled, in the leading order of the spin-orbit interaction, are $S$ and $T_+$
(See appendix B). Although there is no analytical form for this relaxation rate, we have found numerical values showing very slow
leakage rate ($0.001$ s$^{-1}$ or slower when $a,b \ll E_Z - J$) for magnetic fields of interest between 0.1 - 1 T.

In conclusion, we have derived the most general form of the EDSR Hamiltonian for an elliptic quantum dot, in the presence of  the (linear)
spin-orbit interaction. We find a nonlinear $B$-field dependence for the Rabi frequency of the resulting single-spin EDSR. We also propose
electrical manipulation of two-electron-spin states via inter-dot modulation. We find that the phonon-induced triplet-singlet relaxation rates
are typically small, so that coherent Rabi oscillations can, in principle, be observed with current experimental setups in DQD spin-orbit
qubits.

We thank financial support from NSA/LPS through US ARO, and DARPA QuEST through AFOSR.

\appendix

\section{The renormalized g-tensor for confined electrons}

In this subsection, we calculate the renormalization of the confined electron g-factor in the presence of an applied magnetic field. Through the
Schrieffer-Wolff transformation, the effective Hamiltonian (up to second order in spin-orbit interaction and to all orders in Zeeman coupling)
is [5,12]
\begin{eqnarray}
H_{eff} &=& H_d + H_Z + \Delta H, \\
\Delta H &=& \frac{1}{2}[S,H_{so}],
 \end{eqnarray}
where the transformation matrix $S$ is introduced in the main text of the paper. By calculating the above commutator, one can obtain the
(second-order) corrections to the Hamiltonian and the electron spin g-tensor.  The resulting Hamiltonian has a rather complicated form, thus we
only present two special cases of particular interest:

1. {\it B in-plane and along the x axis}
\begin{widetext}
\begin{eqnarray}
\Delta H_{\parallel} &=& - \frac{\hbar^2}{2m} \left(\frac{1}{\lambda_+^2}+\frac{1}{\lambda_-^2} \right)
+ \frac{\hbar^2 \tilde \beta_2}{2m \lambda_+} + \frac{\hbar \, \beta_1}{m \lambda_+}  p_x^2 \, \sigma_x
- \frac{\hbar}{2 m \lambda_-} \{ p_x, p_y \} \, \sigma_y 
+ \left[ \frac{\hbar}{m \lambda_+ \lambda_-} (x p_y - y p_x) - \frac{\hbar \, \tilde \beta_2}{m \lambda_-}  x p_y \right] \sigma_z, \;\;\;\;\;\;\;\;\;\;\;\;\;\;\;\;
\end{eqnarray}
\end{widetext}
where $\{U, V\} \equiv UV + VU$ is the anticommutator of $U$ and $V$.  In an in-plane $B$ field, generally all components of the g-tensor is
renormalized, depending on the electron orbital state.  However, for an electron in the ground state, the quantum mechanical average of the
orbital operators multiplying $\sigma_y$ and $\sigma_z$ vanish, and only $\tilde g_{xx}$ differs from the bulk g-factor.

2. {\it B perpendicular to the x-y plane}
\begin{widetext}
\begin{eqnarray}
 \Delta H_{\perp} &=&  - \frac{\hbar^2}{2m} \left(\frac{1}{\lambda_+^2}+\frac{1}{\lambda_-^2}
 + \frac{\alpha_2}{\lambda_+} -\frac{\beta_2}{\lambda_-} - \frac{\tilde \alpha_2}{\lambda_-}
 - \frac{\tilde \beta_2}{\lambda_+} \right) - \frac{\hbar^2 \omega_c}{2} \left(\frac{\alpha_1}{\lambda_+}
 +\frac{\beta_1}{\lambda_-} + \frac{\tilde \alpha_1}{\lambda_-} - \frac{\tilde \beta_1}{\lambda_+} \right) \nonumber \\
 && +\frac{\hbar}{m}\left[\frac{1}{\lambda_+ \lambda_-}(xp_y - y p_x) + \frac{1}{\lambda_+}\left(\tilde \alpha_1
+ \beta_1\right) p_x^2 +
  \frac{1}{\lambda_-}\left( \alpha_1 - \tilde\beta_1\right) p_y^2 + \frac{1}{\lambda_-}\left(\alpha_2 - \tilde \beta_2 \right)
  x p_y 
+  \frac{1}{\lambda_+}\left(\tilde \alpha_2 + \beta_2 \right) y p_x \right] \sigma_z \;\;\;\;\;\;\;\;\;\;
\end{eqnarray}
\end{widetext}
In this case, only $g_{zz}$ is renormalized, for any electronic orbital state.

\section{The leakage rate out of the singlet-triplet subspace}

\begin{figure}
\begin{center}
\includegraphics[angle=0,width=0.5\textwidth]{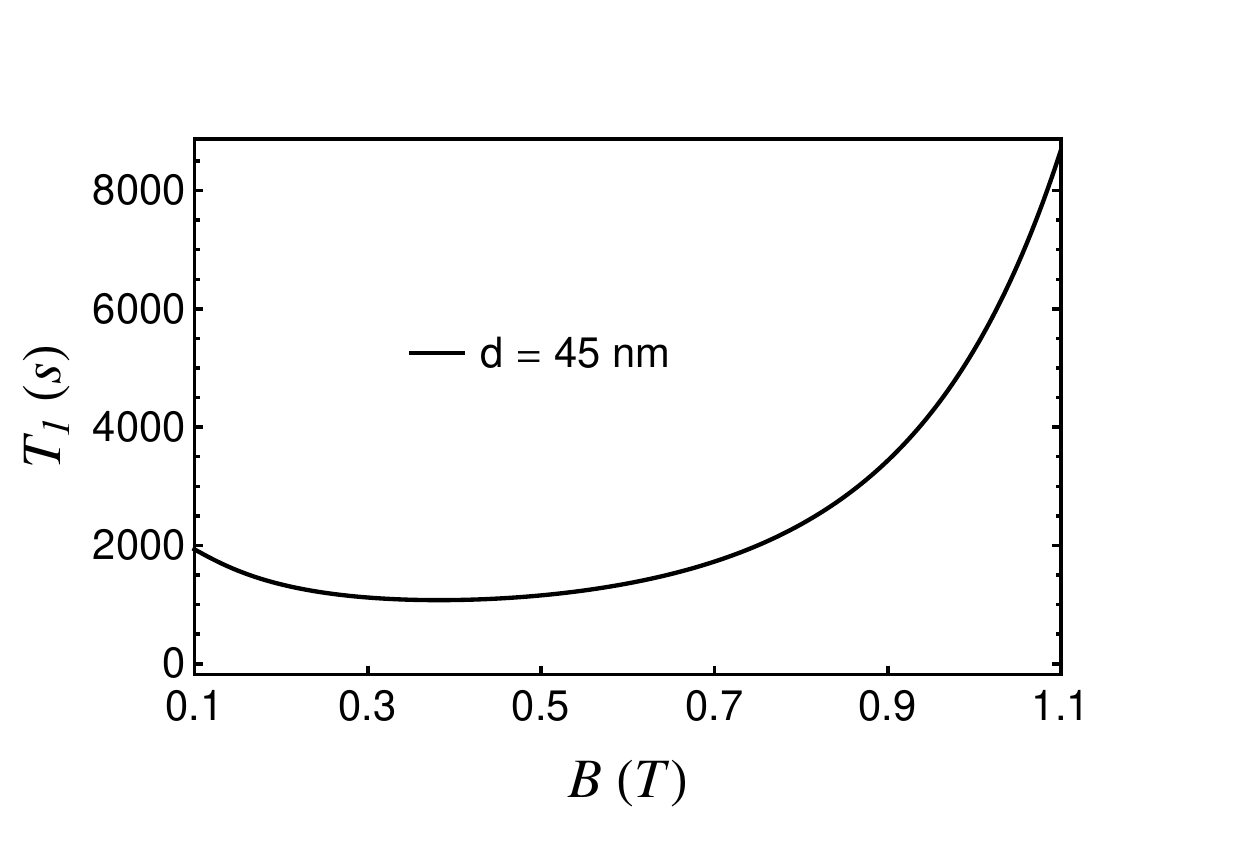}
\caption{\small The singlet-triplet ($S \rightarrow T_+$) relaxation time, for a DQD along the $x$ axis, as a function of applied
(perpendicular) magnetic field.  The material parameters are chosen for GaAs quantum dots with $\lambda_{so} = 5 \; \mu$m,  $l= 30$ nm (at
$B=0$), $e_{14} = 1.4 \times 10^9$ V/m, $\kappa=13.1$, $\rho = 5.3 \times 10^3$ kg/m$^3$, and $s_t = 2.5 \times 10^3$ m/s.  }
\label{Relaxation-supp}
\end{center}
\end{figure}

Here we calculate the two-electron spin relaxation rate for a double dot in a {\it perpendicular} magnetic field, for the regime where $a,b \ll
E_Z - J$.  In a perpendicular $B$ field, there is no coupling between $S - T_0$, in the leading order of the spin-orbit interaction.  The only
leakage channel is $S \rightarrow T_+$ [see Eq. (12) in the text],
\begin{widetext}
\begin{eqnarray}
\Gamma_{S \rightarrow  T_+} &=& \frac{8 (e e_{14})^2}{\pi \kappa^2 \rho \hbar^2 s_t^3} (n_{q_0}+1)
\cdot e^ {-\frac{d^2 (l^4+4l_B^4)}{l^2 l_B^4} } \cdot \frac{  (a_x - b_y)^2 + (a_y - b_x)^2  }{E_Z - J}
\cdot \{ g (d,q_0) +\frac{f(r)}{8} \}, \label{rate-perp} \;\;\; \\
\mbox{\boldmath $a$} &=& \frac{\mu B_z}{2} \left( -\frac{d_x}{\lambda_+}  \, \hat  x+ \frac{ d_y}{\lambda_-} \, \hat y \right), \\
\mbox{\boldmath $b$} &=& \frac{\mu B_z \, l^2}{4 \,  l_B^2} \left( \frac{d_y}{\lambda_+}  \, \hat  x - \frac{ d_x}{\lambda_-} \,
\hat y \right), \\
g (d, q_0) &=& \int_0^\pi d \theta \sin^3\theta   \cos^2\theta \;  e^ {-l^2 q_0^2 \sin^2\theta /2 }
\; I_0 (\frac{2 l^2 d q_0}{l_B^2} \sin\theta) \nonumber\\
&+& \frac {l_B^4}{4 l^4 d^2 q_0^2} \int_0^\pi d \theta \sin^3\theta  \;  e^ {-l^2 q_0^2 \sin^2\theta /2 }
 \; \left\{  I_2 (\frac{2 l^2 d q_0}{l_B^2} \sin\theta) + \frac{2 l^2 d q_0}{l_B^2} \sin\theta \cdot I_3
 (\frac{2 l^2 d q_0}{l_B^2} \sin\theta) \right\},  \nonumber
\end{eqnarray}
\end{widetext}
where $q_0 = \sqrt 2 \; r/l$, $r \equiv l(E_Z -J)/(\sqrt {2} \hbar s_t)$, $l = l_B/\sqrt[4]{\frac{1}{4}+\frac{\omega_0^2}{\omega_c^2}}$, $l_B
=\sqrt{\frac{\hbar c}{e B}}$, and $\omega_c = \frac{e B}{m c}$.  The function $f(x)$ is introduced in the main text of the paper [see Eq.~(14)],
and $I_n(x)$ are the modified Bessel functions of the $n$-th kind. Note that the rate in Eq.~(\ref{rate-perp}) has a geometrical dependence on
the orientation of the setup with respect to the crystallographic axes. In Fig.~\ref{Relaxation-supp}, we have plotted this leakage rate as a
function of the magnetic field for a DQD along the $x$ axis. We find that the rates are typically very small and much slower than the
single-spin (and the hyperfine-induced two-spin) relaxation rates.

\end{document}